\documentclass[aps,showpacs]{revtex4}
\usepackage{graphics}
\usepackage{latexsym}
\usepackage{graphicx,epsfig}
\usepackage{psfrag}
\newcommand{\be}{\begin{equation}}
\newcommand{\ee}{\end{equation}}
\def\bq{\begin{eqnarray}}
\def\eq{\end{eqnarray}}

\begin{document}
%
\large  
\title{SOLUTIONS OF CERTAIN TYPES OF LINEAR AND NONLINEAR DIFFUSION-REACTION EQUATIONS IN ONE DIMENSION}
\pacs{ 05.45.Yv ; 02.30.Ik ; 02.30.Jr }
\author{R. S. Kaushal  \\ Department of Physics, Ramjas College (University Enclave), \\University of Delhi, Delhi-110007, India \\ and \\Department of Physics \& Astrophysics, University of Delhi,\\ Delhi-110007, India}

\email{rkaushal@physics.du.ac.in}

\begin{abstract}
With a view to having further insight into the mathematical content of the non-Hermitian Hamiltonian associaterd with the diffusion-reaction (D-R) equation in one dimension, we investigate (a) the solitary wave solutions of certain types of its nonlinear versions, and (b) the problem of real eigenvalue spectrum associated with its linear version or with this class of non-Hermitian Hamiltonians. For the case (a) we use the standard techniques to handle the quadratic and cubic nonlinearities in the D-R equation whereas for the case (b) a newly proposed method, based on an extended complex phase space, is employed. For a particular class of solutions, an Ermakov system of equations is also found for the linear case. Further, corresponding to the 'classical' version of the above one-dimensional complex Hamiltonian, an equivalent integrable system of two, two-dimensional real Hamiltonians is suggested.
\end{abstract}

\maketitle

\newpage

\section{Introduction}
In mathematical sciences some equations are highly privileged in the sense that their incarnations in analogous forms[1], at times of course with duifferent meanings of the underlying symbols, explain altogether different phenomena in Nature. One such equation, besides the equation of continuity, is the diffusion-reaction (D-R) equation which has offered explanation of many phenomena lying in the domains of not only physics and chemistry but also biology and (now) perhaps social and economic sciences. While the linear version of the D-R equation under its various names such as heat equation, Fokker-Planck equation, Schrodinger-like equation, etc. has been studied very extensively in different contexts, the study of its nonlinear versions and sometimes in higher dimensions have also evoked considerable interest in recent years[2-8].

\par
	Before proceeding further some important remarks about the mathematical contents of the D-R and Schrodinger equations are worth making. The linear version of the D-R equation, namely
 
\begin{equation}
-D\nabla^2C(\bf x,t)+ \bf v.\bf\nabla C(\bf x,t)+ U(\bf x)C(\bf x,t) = -\frac{\partial C(\bf x,t)} {\partial t},
\end{equation}
can be compared with the time-dependent Schrodinger equation

\begin{equation}
  \frac{-1}{2}\nabla^2\psi(\bf x,t)+ V(\bf x)\psi(\bf x,t) = i\frac{\partial \psi(\bf x,t)}{\partial t},
\end{equation}
where $\hbar$=$m$= 1. In eq.(1) $D$ is the diffusion coefficient and the velocity $\bf v$ in general is a function of $\bf x$ and $t$. In the prsent work, however, we shall consider $\bf v$ as a constant, independent of both space and time. Note that in mathematical literature eqs. (1) and (2) are classified differently in view of the different long-time behaviour of their solutions. As a matter of fact, in the limit $t\rightarrow \infty $, while the complete solution of (1) vanishes, the solution of (2) remains periodic in time. The presence of the velocity-term in (1) further makes it distinct from (2). This term, in fact, makes the corresponding 'Hamiltonian' non-Hermitian and as a result the problem of reality of eigenvalues of such a Hamiltonian becomes of considerable interest in view of some recent studies[9-11] in this direction. We shall address some of these issues in the present work. Lastly, while eq.(2) in quantum mechanics is set on the basis of some physical requirements with a rich physics content in it, eq.(1), on the other hand, is just a classical one like any other partial differential equation in mathematics, of course with different contextual meanings of various symbols.

\par
	In the present work, we shall investigate the solutions of one-dimensional linear D-R equation,

\begin{equation}
 C_t + v C_x = D C_{xx} - V(x)C,
\end{equation}
and some of its nonlinear versions, namely

\begin{equation}
  C_t + v C_x = D C_{xx} + [a + U(x)]C - b |C| C,
\end {equation}
and

\begin{equation} 
 C_t + v C_x = D C_{xx} + [a + U(x)]C - b |C|^2 C, 
\end{equation}
where $a$ and $b$ are real constants. We were motivated to study these equations from the recent works of Moiseyev and Gluck[5] and of Nelson and Shnerb[3]. These authors study the three-dimensional version of these equations with reference to the delocalization problem in population biology- a feature of the non-Hermitian character of the Hamiltonian associated with these equations. No doubt, the present study of one-dimensional systems will have limited scope as far as the applications to physical problems are concerned, nevertheless it will provide some clue towards a better understanding of these 'real-world' problems in mathematical terms. While the eq. (4) can be considered as a particular type of generalization of the Malthus-Verhulst growth model[6] in the study of biological systems, eq.(5) is the analogue of nonlinear Schrodinger equation. These cases, to the best of our knowledge, have not been studied earlier. In fact, eq.(4) is a slightly different version of the equation studied by Nelson and Shnerb[3] in the sense that the quadratic nonlinearity now appears as $|C|C$ in stead of $C^2$.

\par
	With regard  to eq.(3), we recast it in the form

\begin{equation}
  {\it H} C(x,t)  = -\frac{\partial C(x,t)}{\partial t},
\end{equation}

where ${\it H}$, a non-Hermitian 'Hamiltonian' operator, is given by[5]

 \[{\it H} = -D \frac{\partial^2}{\partial x^2} + v \frac{\partial}{\partial x} + V(x) .\]
To be more specific, one writes the solution of (3) as $C(x,t)= \psi(x).\tau(t)$ and uses the method of separation of variables to obtain  the eigenvalue equations,

\begin{equation}
\mathcal{H}\psi(x) = \lambda \psi(x),
\end{equation}
with

\begin{equation}
 \mathcal{H} = -D \frac{d^2}{dx^2} + v \frac{d}{dx} + V(x),
\end{equation}
and $(\frac{d\tau}{dt}) = -\lambda \tau$. The solution of the latter equation implies an exponential decay of the complete solution with time. Note that for a complex eigenvalue $\lambda$ there is a possibility of retaining the periodic behaviour of solutions with time- a feature built already in the Schrodinger eq.(2). Further, we write the 'classical' analogue (using $p = -i\frac{d}{dx}$) of $\cal{H}$ as

\begin{equation}   
 H(x,p) = D p^2 + i vp + V(x).
\end{equation}
\par
	Note that for a symmetric potential the non-Hermitian Hamiltonian (9) is symmetric under parity ($x\rightarrow -x$) operation ($P$) and it is not symmetric under time-reversal ($t\rightarrow -t,i\rightarrow -i$) operation ($T$) or under the combined operations of parity and time reversal ($PT$). Therefore, in the light of the conjecture/prescription suggested by Bender et al[9] and used by others[10] eq. (7) can not admit real eigenvalues, as $\cal{H}$ is not a $PT$-symmetric Hamiltonian though it is non-Hermitian. For the case of a symmetric potential $V(x)$, we shall investigate this problem of reality of eigenvalues here within the framework of a complex phase space approach proposed[12] and used recently[13] for a variety of complex potentials. In this approach, the real $(x,p)$phase plane is extended to a complex phase space characterized by writing $x$ and $p$ as
\begin{equation}
 x = x_1 + ip_2 ;          p = p_1 + ix_2,
\end{equation}
where $(x_1,p_1)$ and $(x_2,p_2)$ turn[15] out to be the canonical pairs in an equivalent real two-dimensional space. The arrangement of the paper is as follows:

\par
	In Sect.2 we investgate the solitary wave solutions of the nonlinear eqs.(4) and (5). We study the eigenvalue problem associated with the non-Hermitian operator (8) in Sect.3. In Sect.4, we highlight some other, so far unexplored, mathematical features of the Hamiltonian (8) (or (9)). In particular, for the linear case, in analogy with the Schrodinger equation[16] an Ermakov 
system[17] of equations is derived in this section. Also, corresponding to the one-dimensional complex Hamiltonian (9), an equivalent system of two, two-dimensional real Hamiltonians is obtained for an analytic potential function $V(x)$. Finally, concluding remarks are made in Sect. 5.

\section{Solitary Wave Solutions of the Nonlinear D-R Equations}
In this section we obtain solitary wave solutions of the nonlinear eqs.(4) and (5).

\subsection{Solution of eq.(4)}
For the traveling wave solution of eq.(4) we consider the case when the random potential $U(x)$ is constant, say $U_0$, and for the solution $C(x,t)$ we make an ansatz

\begin{equation}
 C(x,t) = r(\xi). exp[i\theta(\xi)+\delta t],
\end{equation}
where $\xi = x- w t$, and $\delta$ and $w$ are arbitrary real constants to be  determined later. Using (11) in (4) and then separating the real and imaginary parts of the resultant expression, one obtains

\begin{equation}
 (v- w)r'+ r\delta = D r'' - D r\theta '^2 + (a+U_0) r - b r^2,
\end{equation}
\begin{equation}
 (v-w) r \theta' = D (2 r' \theta' + r \theta'').
\end{equation}
Here, the primes indicate the derivatives with respect to the variable $\xi$. Next we look for the solutions of these coupled total differential equations in $r(\xi)$ and $\theta (\xi)$ under some simplifying assumtions.

\par
	After defining $y = r^2\theta'$ for the right hand side, eq.(13) can easily be recast in the form, $y' = ((v- w)/D)y$, which admits a solution $y = y_0 \exp[((v- w)/D)\xi]$, or

\begin{equation}
 \theta' = (k/r^2).\exp[((v- w)/D)\xi],
\end{equation}
where $y_0$ ( or $k$) is the integration constant. For simplicity, we concentrate here on the case when $ w =v$, i.e., when the parameter $w$ in the ansatz (11) becomes the convective velocity $v$ of the system. This leads to $\theta' = k/r^2$.

\par
	For the above choices, eq. (12) can be expressed as

 \[ r'' = \frac{k^2}{r^3} + \frac{1}{4}B r + \frac{3}{8}A r^2 , \] 

which can be easily integrated to give

\begin{equation}
 (r')^2 = -\frac{k^2}{r^2} + \frac{1}{4}B r^2 + \frac{1}{4}A r^3 + \frac{k_1}{4},
\end{equation}
where $(k_1/4)$ is the constant of integration and $B= 4(\delta -a- U_0)/D$, $A=(8b/3D)$. Alternatively, by defining $S= r^2$ we write eq.(15) in terms of the variable $S$ as

\begin{equation}
 (S')^2 = A \sqrt{S^5} + B S^2 + k_1 S - 4k^2.
\end{equation}
For the solitary wave solutions, we set $k_1 = k =0$, therby reducing eq.(16) to the form $S' = S(A \sqrt{S} + B)^{1/2}$ or eq. (15) to

\begin{equation}
 r' = (1/2) r \sqrt{A r + B},
\end{equation}
which can be integrated to give[18] $r(\xi)$ as
\[ r(\xi) =-(B/A).\sec^2(\frac{\sqrt{-B}}{4} \xi +\xi_0) , \] 
for $B<0$, and
 
 \[r(\xi) = (B/A). cosec h^2(\frac{\sqrt{B}}{4} \xi +\xi_0) , \] 
for $B>0$. Correspondingly, the solution of $\theta' = 0$ equation is taken as $\theta = \rm constant$ (say $\theta_0$). Finally, the solutions of (4) in view of (11) becomes

\begin{equation}
 C(x,t) =-(B/A).\exp(i\theta_0 +\delta t).\sec^2(\frac{\sqrt{-B}}{4} \xi +\xi_0),   (B<0),
\end{equation}
and
 
\begin{equation}
 C(x,t) = (B/A).\exp(i\theta_0 +\delta t). cosec h^2(\frac{\sqrt{B}}{4} \xi +\xi_0),   (B>0),
\end{equation}
where $\xi_0 (=x_0-vt_0)$ is the constant of integration and the same can be fixed from the initial conditions.

\subsection{Solution of eq.(5)}
Following the same procedure as for eq.(4) in the above subsection with ansatz (11) , the imaginary part of the resultant expression, in the present case, will yield the same equation as eq.(13). The real part however now becomes

 \[ (v- w)r'+ r\delta = D r'' - D r\theta '^2 + (a+U_0) r - b r^3 .\] 
An equation analogous to eq.(16) for the present case can be derived as

\begin{equation}
  (S')^2 = A S^3 + B S^2 + k_1 S - 4k^2.
\end{equation}
Under the same simplifying assumptions as made in the above subsection for the case of solitary wave solutions, the solution of $\theta' = 0$ equation is again taken as $\theta = \theta_0$. The equation analogous to eq.(17) now becomes

\begin{equation}
 r' = (1/2) r \sqrt{A r^2 + B},
\end{equation}
where $B$ is the same as before but $A=(2b/D)$. Eq.(21) can be solved[18] to give

  \[r(\xi) =\sqrt{-B/A}.\sec(\frac{\sqrt{-B}}{2} \xi +\xi_0),       (B<0),\] 

  \[r(\xi) = \pm \sqrt{B/A}. cosec h (\frac{\sqrt{B}}{2} \xi +\xi_0),    (B>0).\]

Finally, the solitary wave solution of eq.(5) can be written as 
\begin{equation}
  C(x,t)=\sqrt{-B/A}.\exp(i\theta_0 +\delta t).\sec(\frac{\sqrt{-B}}{2} \xi +\xi_0),   (B<0), 
\end{equation}

\begin{equation}
 C(x,t)=\pm \sqrt{B/A}.\exp(i\theta_0 +\delta t).cosec h (\frac{\sqrt{B}}{2} \xi +\xi_0), (B>0).
\end{equation}
where $\xi_0$ is the constant of integration to be determined from the initial conditions.

\section{Eigenvalue Problem Associated with Eq.(7)}
\subsection{General results}
Since the Hamiltonian operator (8) is non-Hermitian, the eigenvalue $\lambda$ in (7) need not be real. Further for a symmetric potential, $\cal{H}$ is also not a $PT$-symmetric (a relaxed case of non-Hermiticity) one and hence may not admit[9] real eigenvalues. Naturally, the conjecture of Bender et al[9] for the reality of eigenvalues of $PT$-symmetric potentials (developed mainly in the context of the Schrodinger equation) is bound to show some limitations in this case. While the concept of pseudo-Hermiticity[11] may be worth attempting for the present case, we have however been pursuing[12,13] an altogether different method to handle the complex Hamiltonian systems. From this point of view our approach is quite general and the concept of an extended complex phase space defined by (10) is used. In what follows we investigate the solution of an analogous D-R equation (7), in the sense that x and p in it are now complex.
\par
	Note that $V(x)$ in (7) in general could be complex just as $x$, $p$ and $\psi$ are. Thus, we write $V(x) = V_r(x_1,p_2) + i V_i(x_1,p_2), \psi(x) = \psi_r(x_1,p_2)+ i \psi_i(x_1,p_2), \lambda = \lambda_r + i \lambda_i $, and

 \[ \frac{d}{dx} = \frac{\partial}{\partial x} - i \frac{\partial}{\partial p_2};    \frac{d^2}{dx^2} = \frac{\partial^2}{\partial x_{1}^{2}} - 2i \frac{\partial^2}{\partial x_1 \partial p_2} - 
\frac{\partial ^2}{\partial p_{2}^{2}} ,\]
and use them in (7) to give
 
\begin{eqnarray}
 -D [\psi_{r,x_1x_1} - 2i \psi_{r,x_1p_2} - \psi_{r,p_2p_2} + i\psi_{i,x_1x_1} + 2\psi_{i,x_1p_2} - i\psi_{i,p_2p_2}] + v [\psi_{r,x_1} - i\psi_{r,p_2} + i\psi_{i,x_1} + \psi_{i,p_2}] \nonumber\\
+ V_{r}\psi_{r} - V_{i}\psi_{i} + i V_{r}\psi_{i} +i V_{i}\psi_{r} = \lambda_{r}\psi_{r} -  \lambda_{i}\psi_{i} + i\lambda_{i}\psi_{r} +i\lambda_{r}\psi_{i}.
\end{eqnarray}
Now we equate the real and imaginary part of this expression separately to zero. This leads to the following pair of coupled partial differential equation in $\psi_{r}$ and $\psi_{i}$ :

 \[-D [\psi_{r,x_1x_1} - \psi_{r,p_2p_2} + 2\psi_{i,x_1p_2}] + v [\psi_{r,x_1} + \psi_{i,p_2}] + V_{r}\psi_{r} - V_{i}\psi_{i} = \lambda_{r}\psi_{r} - \lambda_{i}\psi_{i} ,\]

 \[-D [-2 \psi_{r,x_1p_2} + \psi_{i,x_1x_1} - \psi_{i,p_2p_2}] + v [- \psi_{r,p_2}+ \psi_{i,x_1}]+ 
V_{r}\psi_{i} + V_{i}\psi_{r} = \lambda_{i}\psi_{r} + \lambda_{r}\psi_{i} .\]

Further use of the Cauchy-Riemann conditions for the analyticity of $\psi (x)$, viz.,

\begin{equation}
 \psi_{r,x_1} = \psi_{i,p_2} ;    \psi_{r,p_2} = -\psi_{i,x_1} ,
\end{equation}
leads to simpler forms of these equations, namely
\begin{equation}
 -4D \psi_{r,x_1x_1} + 2v \psi_{r,x_1} + V_{r}\psi_{r} - V_{i}\psi_{i} = \lambda_{r}\psi_{r} - \lambda_{i}\psi_{i},
\end{equation}
\begin{equation}
 -4D \psi_{i,x_1x_1} + 2v \psi_{i,x_1} + V_{r}\psi_{i} + V_{i}\psi_{r} = \lambda_{i}\psi_{r} + \lambda_{r}\psi{i}.
\end{equation}
\par
	As for the case of Schrodinger equation[12], we make an ansatz here for the solution,viz.,
\begin{equation}
 \psi (x) \equiv \psi_{r}+i\psi_{i}= e^{g(x)} ; g(x)= g_{r}(x_1,p_2)+i g_{i}(x_1,p_2) ,
\end{equation}
which gives $\psi_{r}(x_1,p_2)= e^{g_{r}}\cos{g_{i}}$,$\psi_{i}(x_1,p_2)= e^{g_{r}}\sin{g_{i}}$ or $g_{r}=(1/2)\ln(\psi_{i}^2 + \psi_{r}^2)$, $g_{i}= tan^{-1}(\psi_{i}/\psi_{r})$ . Now, after using (28) in (26) and rationalizing the resultant expression with respect to the orthogonal functions $\cos(g_{i})$ and $\sin(g_{i})$, we obtain the following pair of coupled partial differential equations:
\begin{equation}
 g_{r,x_1x_1} + (g_{r,x_1})^2 - (g_{i,x_1})^2 - (v/2D) g_{r,x_1} + (1/4D)(\lambda_{r}-V_{r}) = 0,
\end{equation} 
\begin{equation}
 g_{i,x_1x_1} + 2 g_{r,x_1}.g_{i,x_1} - (v/2D) g_{i,x_1} + (1/4D)(\lambda_{i}-V_{i}) = 0.
\end{equation} 
Interestingly, the same set of equations is also arrived at if one rationalizes the eq.(27) using (28). Thus, for a given complex potential, eqs.(29) and (30) in which the original ansatz for $\psi (x)$ is now transcribed into that for $g(x)$ can be solved to obtain the real and imaginary parts of the eigenvalue $\lambda$. Further, we shall consider the two situations- one when the parameter(s) of the potential $V(x)$ are real or the other when they are complex. In the following we demonstrate the applications of these general results to the case of a simple harmonic oscillator corresponding to these two situations.

\subsection{Applications to complex harmonic oscillator potential}

We first consider the case of a complex harmonic oscillator potential,
\begin{equation}
 V(x) = a x^2
\end{equation}
where $a$ is real. Using (10), we write $V_{r}(x_1,p_2)= a (x_1^2-p_2^2)$, $V_{i}(x_1,p_2)= 2 a x_1 p_2 $ and make the following ansatz for $g_{r}$ and $g_{i}$ which is consistent with the Cauchy-Riemann conditions:
\begin{equation}
 g_{r}(x_1,p_2)=\alpha_1 x_1 +\alpha_2 p_2 +\alpha_{20} (x_1^2 -p_2^2) +\alpha_{11} x_1 p_2 ; g_{i}(x_1,p_2)=-\alpha_2 x_1 +\alpha_1 p_2 - \frac{1}{2}\alpha_{11} (x_1^2 -p_2^2) +2\alpha_{20}x_1 p_2,
\end{equation}
where $\alpha_{i}$, $\alpha_{ij}$'s are real constants to be determined later. Substitution of (32) in (29) and subsequent rationalization of the resultant expression with respect to the powers of $x_1$, $p_2$ and $(x_1p_2)$ yields the following relations:

\begin{equation}
 2\alpha_{20} + \alpha_1^2 - \alpha_2^2 - (v/2D)\alpha_1 + (1/4D)\lambda_{r} = 0,
\end{equation} 
\begin{equation}
 4\alpha_1 \alpha_{20} - 2\alpha_2 \alpha_{11} - (v/D)\alpha_{20} = 0,
\end{equation} 
\begin{equation}
 4\alpha_2 \alpha_{20} + 2\alpha_1 \alpha_{11} - (v/2D)\alpha_{11} = 0,
\end{equation} 
\begin{equation}
 4\alpha_{20} \alpha_{11} + 4\alpha_{20}\alpha_{11} = 0,
\end{equation} 
\begin{equation}
 4\alpha_{20}^2 - \alpha_{11}^2 - (a/4D) = 0,
\label{eqn:alp0} \end{equation} 
\begin{equation}
   \alpha_{11}^2 - 4\alpha_{20}^2 + (a/4D) = 0 .
\end{equation} 
Note that while eqs. (37) and (38) turn out to be identical, eq.(36) gives $\alpha_{11}\alpha_{20}=0$ which implies that either $\alpha_{20}=0$ or $\alpha_{11}=0$ (of course both of them can not be set equal to zero). Again $\alpha_{20}=0$ is ruled out as it leads to imaginary $\alpha_{11}$ from (37). Therefore, $\alpha_{11}=0$ and $\alpha_{20}=\pm \sqrt{a/16D}$ from (37). Eqs.(34), (35) and (33) immediately give $\alpha_{1}=(v/4D)$, $\alpha_{2}=0$ , and the real part of $\lambda$ as

\begin{equation}
 \lambda_{r} = \mp 2\sqrt{aD} + (v^2/4D) .
\end{equation}

\par
	Similarly, if one rationalizes (30) using (32) the same set of equations and hence the same values (as above) of the unknown parameters in the ansatz (32) are obtained except for the equation involving $\lambda_{i}$. The imaginary part,$\lambda_{i}$, of $\lambda$ now turns out to be zero, which implies only the real eigenvalue for the potential (31). Thus, for the potential (31), we have
\begin{equation}
 \lambda_{r} = \mp 2\sqrt{aD} + (v^2/4D)  ; \lambda_{i} = 0 ,
\end{equation}
 \[g_{r}(x_1,p_2) = (v/4D)x_1 \pm \sqrt{a/16D} (x_1^2 - p_2^2); g_{i}(x_1,p_2) = (v/4D)p_2 \pm \sqrt{a/4D}x_1 p_2 \]
or equivalently,
\begin{equation}
 \psi (x) = exp[(v/4D) x \pm \sqrt{a/16D} x^2].
\end{equation}
\par
	Next, we consider the case when the parameter $a$ of the potential (31) is complex, $a = a_{r}+i a_{i}$. In that case we have
\begin{equation}
 V_{r}(x_1,p_2)= a_{r}(x_1^2 - p_2^2) - 2a_{i}x_1 p_2 ;  V_{i}(x_1,p_2)= a_{i}(x_1^2 - p_2^2) - 2a_{r}x_1 p_2 .
\end{equation}
As before, the use of ansatz (32) in (29) now yields the following set of equations after the rationalization of the resultant exptression:
\begin{equation}
 2\alpha_{20} + \alpha_1^2 - \alpha_2^2 - (v/2D)\alpha_1 + (1/4D)\lambda_{r} = 0,
\end{equation}  
\begin{equation}
 4\alpha_2 \alpha_{20} + 2\alpha_1 \alpha_{11} - (v/2D)\alpha_{11} = 0,
\end{equation} 
\begin{equation}
 4\alpha_1 \alpha_{20} - 2\alpha_2 \alpha_{11} - (v/D)\alpha_{20} = 0,
\end{equation} 
\begin{equation}
 4\alpha_{20} \alpha_{11} + 4\alpha_{20}\alpha_{11} + (a_{i}/2D) = 0,
\end{equation} 
\begin{equation}
 4\alpha_{20}^2 - \alpha_{11}^2 - (a_{r}/4D) = 0,
\end{equation} 
\begin{equation}
   \alpha_{11}^2 - 4\alpha_{20}^2 + (a_{r}/4D) = 0 .
\end{equation}
Eqs.(44)-(47) can be solved for the real parameters $\alpha_1$, $\alpha_2$, $\alpha_{20}$ and $\alpha_{11}$ in the ansatz (32) and in terms of the complex parameter $a$ of the potential (31) to give
\[ \alpha_1 = (v/4D) ; \alpha_2 = 0; \alpha_{20}= a_{+}/4\sqrt{2D}; \alpha_{11}= -a_{-}/2\sqrt{2D} ,\]
where $a_{\pm}=\sqrt{|a| \pm a_{r}}$. Using these results in (43), one obtains $\lambda_{r}$\ as
\begin{equation}
 \lambda_{r} = \mp \sqrt{2D} a_{+} + (v^2/4D) .
\end{equation}
Similarly, the rationalization of eq.(30) using (32) and (42) will give the imaginary part, $\lambda_{i}$, of the eigenvalue in a straightforward manner,viz.,
\begin{equation}
 \lambda_{i} = \pm \sqrt{2D}  a_{-} .
\end{equation}
Finally, the eigenfunction $\psi (x)$ in terms of $g_{r}(x_1,p_2)$ and $g_{i}(x_1,p_2)$ turns out to be
 \[g_{r}(x_1,p_2) = (v/4D)x_1 + (a_{+}/4\sqrt{2D})(x_1^2 - p_2^2) - (a_{-}/2\sqrt{2D})x_1 p_2 ;\]  \[g_{i}(x_1,p_2) = (v/4D)p_2 + (a_{-}/4\sqrt{2D})(x_1^2 - p_2^2) + (a_{+}/2\sqrt{2D}) x_1 p_2 ,\]
or equivalently,
\begin{equation}
 \psi (x) = exp[(v/4D) x + (1/4\sqrt{2D})(a_{+} + ia_{-}) x^2].
\end{equation}

\par
	Before we highlight two special cases of these general results it can be seen that the eigenvalues for the non-Hermitian operator (8) are real as long as the potential parameter $\it{a}$
is real. Once $\it {a}$ becomes complex, then the complexicity of the eigenvalue may arise as a result of $a_{i}\neq 0$.
\par
	Note that for the case when $D = (1/2)$ and $v = 0$, above results for the complex harmonic oscillator trivially reduce to those derived by solving the Schrodinger equation in extended complex phase space (cf. Ref.(13), Sect. 3). Further, if we set $a_{i}=0$, $a_{r}=a=|a|$ (or $a_{-}=0$ , $a_{+}=\sqrt{2a}$), then it is not difficult to see that results (49)-(51) for the complex coupling reduce to that for the case of real coupling (cf. eqs.(40) and (41)).
\par
	Mainly for the sake of a comparision we present here the results for the case of real harmonic oscillator, $V(x)= ax^2$, obtained by solving (7) in a real phase plane. Again using the ansatz $\psi (x)= exp[g(x)]$, with $g(x)=\alpha x^2 + \beta x$ for the solution, it is not difficult to arrive at the following results for $\lambda$ and $\psi (x)$ :
  
\begin{equation}
 \lambda = \mp \sqrt{aD} + (v^2/4D) ,
\end{equation}
\begin{equation}
 \psi (x) = exp[(v/2D) x \pm \sqrt{a/4D} x^2].
\end{equation}
It can be seen that the results (52) and (53) differ slightly from (40) and (41) in terms of numerical factors. This is maily because of the requirement of analyticity of $\psi (x)$ used in the case of complex phase space.

\section{Other Features of the Diffusion Hamiltonian (8)}
In this section we discuss some previously unnoticed mathematical features of the D-R Hamiltonian (8) (or its 'classical' version (9)). In the first part we derive an Ermakov system of equations for (8) and in the second, following the prescription advanced in our earlier work[15], we derive an equivalent integrable system of two, two-dimensional real Hamiltonians corresponding to the complex Hamiltonian (8).

\subsection{Ermakov system of equations and space invariant}
About 125 years ago, Ermakov for the first time demonstrated[17] the linkage between the solutions of certain type of differential equations via an integral invariant. This latter construct, now termed as 'Ermakov' (or 'Lewis') invariant in the context of classical mechanics, has played[19]  
an important role in the study of time-dependent systems and in the quantum[16] and other contexts[21,22] several new interpretations of this mathematical construct have been sought. Here, since the variable x characterizes the space dimension, we shall call this construct as the 'space invariant'.

\par
	For the present purpose, we rewrite eq.(7) as
\begin{equation}
 \psi'' - \gamma \psi' + q^2(x) \psi = 0,
\end{equation}
where $\gamma = (v/D)$, $q^2(x) = (\lambda - V(x))/D$ , and look for its solutions in the form (ansatz)
\begin{equation}
 \psi (x) = \phi (x) exp[ i f(x)].
\end{equation}
Now, after using (55) in (54) and equating the real and imaginary parts separately to zero in the resultant expression, we obtain
\begin{equation}
 \phi''- (f')^2 \phi -\gamma \phi' + q^2 \phi = 0,
\end{equation}
\begin{equation}
 f''\phi + 2 f'\phi' - \gamma \phi f' = 0.
\end{equation}

\par
	As before, eq.(57) after defining $y = \phi^2 f'$, can easily be recast in the form $y' = \gamma y$, whose solution now becomes $y = \kappa .exp(\gamma x)$, or
\begin{equation}
 f' = \kappa  e^{\gamma x}/ \phi^2 ,
\end{equation}
where $\kappa $ is the constant of integration. While the integration of (58) in the form
\begin{equation}
 f(x) = \kappa \int^{x} (e^{\gamma x'}/\phi^2 (x')) dx' ,
\end{equation}
suggests a phase-amplitude connection in the present case, its use in (56) leads to a nonlinear differential equation,
\begin{equation}
 \phi'' - \gamma \phi' + q^2 \phi = \kappa^2 e^{2\gamma x}/\phi ^3 .
\end{equation}

\par
	In order to derive the space invariant, multiply eq.(54) by $\phi$ and eq.(60) by $\psi$ and subtract the latter to give,
\begin{equation}
 (\phi \psi'' -\phi''\psi) + \gamma (\psi \phi' -\psi'\phi) = -\kappa^2 e^{2\gamma x}\psi /\phi ^3.
\end{equation}
This expression, after using $2(\phi \psi' -\phi'\psi) $ as the integrating factor, can immediately be integrated to give the space invariant, $K$, in the form
\begin{equation}
 K = \kappa^2 (\psi/\phi)^2 + e^{-2\gamma x}(\phi \psi' -\phi'\psi)^2 .
\end{equation}
Note that the structure (62) satisfies $(dK/dx) = 0 $, and hence termed as 'space invariant'. Clearly, it is a manifestation of the phase-amplitude connection (59). Although, $K$ as such appears to be independent of the form of $V(x)$ but the fact is that the role $V(x)$ enters through (60) or for that matter via (59). The system of eqs. (54), (60) and (62) constitute an Ermakov system.

\subsection{An equivalent integrable system in two real dimensions}
In our earlier[15] work, we have exploited the analyticity property of the one-dimensional complex Hamiltonian $H(x,p)$ by writing it in the form $H(x,p) = H_{1}(x_1,x_2,p_1,p_2) + i H_{2}(x_1,x_2,p_1,p_2)$ (cf. eq.(10)). This has resulted into a new class of two-dimensional integrable real Hamiltonian systems described by $H_{1}(x_1,x_2,p_1,p_2)$ or by $ H_{2}(x_1,x_2,p_1,p_2)$ for certain choices of the complex potential $V(x)$. Here we shall demonstrate that this prescription also works for the complex Hamiltonian (9) and more so independently of the form of $V(x)$ which is now expressed as $V(x)= V_{r}(x_1,p_2) + i V_{i}(x_1,p_2)$. 

\par 
	Note that the 'classical' version(9) of the Hamiltonian (8) is complex even in the real phase plane. In the complex phase space (10), this version of $\cal{H}$ , expressed as $H(x,p) = H_{1}(x_1,x_2,p_1,p_2) + i H_{2}(x_1,x_2,p_1,p_2)$, becomes a function of two complex variables $x$ and $p$ and its real and imaginary parts turn out to be

\begin{equation}
 H_{1}(x_1,x_2,p_1,p_2) = D (p_1^2 - x_2^2) - v x_2 + V_{r}(x_1,p_2) ,
\end{equation}  
\begin{equation}
 H_{2}(x_1,x_2,p_1,p_2) = 2D p_1 x_2 + v p_1 + V_{i}(x_1,p_2) .
\end{equation} 

Next, we compute the Poisson bracket $[H_1,H_2]_{PB}$ from

\begin{equation}
 [H_1,H_2]_{PB} = \frac{\partial H_1}{\partial x_1}.\frac{\partial H_2}{\partial p_1} - \frac{\partial H_1}{\partial p_1}.\frac{\partial H_2}{\partial x_1} + \frac{\partial H_1}{\partial x_2}.\frac{\partial H_2}{\partial p_2} - \frac{\partial H_1}{\partial p_2}.\frac{\partial H_2}{\partial x_2}.
\end{equation}
It can be noticed that for $H_1$ and $H_2$ given by (63) and (64) the poisson bracket (65) vanishes, if and only if the real and imaginary parts of $V(x)$ satisfy the Cauchy-Riemann conditions, viz., $V_{r,x_1} = V_{i,p_2}$ , $V_{i,x_1} = -V_{r,p_2}$. In other words, this implies the analyticity of $V(x)$. Further, vanishing of the Poisson bracket (65) also suggests that $H_1$ and $H_2$ are in involution and independent in the sense[15] that vectors $\nu_{j} = J\nabla_{y} H_{j}(y)$ for $j=1,2$ and $y=x_1,x_2,p_1,p_2 $ turn out to be linearly independent for (63) and (64). Here $J$ is the symplectic unit matrix. It may be mentioned that $H_1$ and $H_2$ as given by (63) and (64) fulfil all the other requirements listed in Ref.(15) for a biharmonic function or for an auto-Backlund transformation and thereby enabling the integrability of the system. 
\par
	For the integrability of a two-dimensional, time independent system one expects the existence of another invariant in addition to the given Hamiltonian of the system. Such a second invariant, if becomes available, helps in quadrature. While the search for such an invariant has been there in the literature for many years and for many systems[19], the same has been constructed only for a few countable ones. In spite of many developments in terms of various methods for this purpose[19], there still remains scarcity of integrable systems. As a matter of fact, in some cases, even if the existence of the invariant is confirmed, its construction becomes a problem. From this point of view, the prescription suggested in Ref.(15) offers some kind of relief in the sense that at least for a certain class of Hamiltonian systems the second invariant becomes automatically available. In the present case, the vanishing of the Poisson bracket (65) not only suggest that $H_1$ and $H_2$ are in involution but also that if $H_1(x_1,x_2,p_1,p_2)$ is the Hamiltonian of the system, then $H_2(x_1,x_2,p_1,p_2)$ is the second invariant of this system or the vice-versa, and thus implying the integrabilty of the system. No doubt the structures of the Hamiltonians, in this case, depart from the conventional ones but the fact is that such structures have also become desirable to understand some newly discovered phenomena in nature (see the citations in Ref.(15)).

\par
In view of the above general discussion, a few remarks about the Hamiltonian corresponding to the case of complex oscillator (cf. Sect. 3(B) ) are in order. Note that the potential $V(x)$ in this case is an analytic function only when the parameter $a$ is real. Corresponding to this case the equivalent integrable system in two real dimensions (cf. eqs. (63) and (64)) turns out to be

\begin{equation}
  H_{1}(x_1,x_2,p_1,p_2) = D (p_1^2 - x_2^2) - v x_2 + a (x_1^2 - p_2^2),
\end{equation}  
\begin{equation}
  H_{2}(x_1,x_2,p_1,p_2) = 2D p_1 x_2 + v p_1 + 2 a x_1 p_2 .
\end{equation} 
These forms of $H_1$ and $H_2$ are worth comparable to any one of the examples studied in Ref.(15). As a matter of fact for the case of complex parameter the potenial $V(x)$ is no longer an analytic function in the spirit of the above Cauchy-Riemann conditions, in stead it becomes the function of two complex variables $a$ and $x$. For the analyticity of such a function one has to use a generalized version of Cauchy-Riemann conditions[23] which, in fact, involve the higher  derivatives of $V_{r}$ and $V_{i}$; whereas the computation of the Poisson bracket (65) requires only the first derivatives of $V_{r}$ and $V_{i}$.

\section{Concluding Discussion}
From the point of view of learning more about the D-R equation, two slightly disconnected aspects of this equation are explored in this work. In the first part, we have studied the solitary wave solutions of the D-R equation with some specific types of quadratic and cubic nonlinearities, which, to the best of our knowledge, have not been investgated earlier in this context. With regard to the application of these results, it concerns the modelling part of the study of a nonlinear phenomenon. In fact, there appear now many situations in the fields of population biology or in different branches of physics and chemistry, where the results obtained here can be useful in offering the alternatives while accounting for the nonlinearity in such studies.

\par
	We have restricted ourselves only to the study of solitary wave solutions of eqs.(4) and (5) by way of ignoring certain terms in eqs. (16) and (20). This is done mainly for simplicity ; otherwise one can as well retain all the terms in these equations and integrate the resultant expressions to obtain the traveling wave solutions of more general type. In fact, it is not difficult to arrive at the cnoidal wave solution in case of eq. (5) for certain choices of the parameters and in analogy with the KdV equation[20].

\par
	In spite of the fact that the Hamiltonian (8) is non-Hermitian, it does not admit complex eigenvalues in general (cf. Sect. 3 ). The complex eigenvalues can, however, be expected if the parameter(s) of the potential  also becomes complex in addition to the phase space. In some sense this will bring the D-R equation closer to the Schrodinger equation in spite of the presence of the velocity-term in the former, particularly with reference to the large-$t$ behaviour of their solutions.

\par
	Two previously unexplored (to the best of our knowledge) aspects of the D-R Hamiltonian (8) are highlighted in the last section. These studies only hint to the richness of the mathematical content present already in the Hamiltonian (8) or in its 'classical' version (9). In analogy with other studies[16,19,21,22], a couple of possible interpretations of the constructed space invariant $K$ (cf. eq.(62)) can be re-emphasized here in the present context. Firstly, since $K$ is a space invariant and involves the solutions of both the eqs. (54) and (60), it can act as a geometric constraint with regard to the validity of these solutions. Secondly, as the invariant $K$ is the manifestation of a particular type of phase-amplitude connection (59) via the nonlinear eq. (60), its existence itself suggests[19,22] a nonlinear superposition principle in which the solution of (60) is expressible in terms of  two linearly independent solutions of (54). Further applications of some of these results to the stellar structure studies are in progress.
\vspace{1in}

\begin{acknowledgements}

A part of this work was done when the author was visiting the Inter-University Centre for Astronomy and Astrophysics (IUCAA), Pune as an Associate. He wishes to thank the Director, Professor N. Dadhich and Professor A. Kembhavi for the facilities, Professors T. Padmanabhan and V. Sahni for several helpful dicussions. Thanks are also due to Drs. D.Parashar and M. Sami for a critical reading of the manuscript.

\end{acknowledgements}

\newpage

{\bf \centerline {REFERENCES}}
\vspace{1cm}
[1]. Radhey Shyam Kaushal, {\sf Structural Analogies in Understanding Nature} (Anamaya Publishers, New Delhi, 2003 ).\\

[2]. N. Hatano and D. R. Nelson, Phys. Rev. Lett. {\bf 77} (1996) 570.\\

[3]. D. R. Nelson and N. M. Shnerb, Phys. Rev. {\bf E58} (1998) 1383, and the references therein, also see, N. M. Shnerb and D.R. Nelson, Phys. Rev. Lett. {\bf 80} (1998) 5172.\\

[4]. J. T. Chalker and Z. Jane Wang, Phys. Rev. Lett. {\bf 79} (1997) 1797.\\

[5]. N. Moiseyeb and M. Gluck, Phys. Rev. {\bf E63} (2001) 041103.\\

[6]. J. D. Murray,  {\sf Mathematical Biology}, (Springer-Verlag, New York, 1993 ).\\

[7]. K. B. Efetov, Phys. Rev. Lett. {\bf 79} (1997) 9630; I. V. Goldsheid and B. A. Khoruzhenko, Phys. Rev. Lett. {\bf80} (1998) 2897.\\

[8]. R. S. Kaushal and D. Parashar, {\sf Advanced Methods of Mathematical Physics} (Narosa Publishing House, New Delhi, CRC Press, USA, Alpha Science Int., London, 2000) Chapter 10, Also see, D.R.King, J. Phys. {\bf A}:Math \& Gen. {\bf 24} (1991) 3213; ibid. {\bf 23} (1990) 3681.\\

[9]. C. M. Bender and S. Boettcher, Phys. Rev. Lett. {\bf 80} (1998) 5243; C.M. Bender, S. Boettcher and P.N. Meisinger, J. Math. Phys. {\bf 40} (1999) 2201, and the references therein.\\

[10]. M. Znojil and G. Lavai, Phys. Lett. {\bf A271} (2000) 327; F.M. Fernandez et al, J. Phys.{\bf A}:Math \& Gen. {\bf 31} (1998)10105; B. Bagchi, F. Cannata and C. Quesne, Phys. Lett. {\bf A269} (2000) 79; Z. Ahmed, Phys. Lett. {\bf A294} (2002) 287, and the citations in Ref.(13) below.\\

[11]. A. Mustafazadeh, J. Math. Phys. {\bf 43} (2002) 2814; 205, and the references therein.\\

[12]. R.S. Kaushal,J. Phys.{\bf A}:Math \& Gen. {\bf34} (2001) L709.\\

[13]. R.S. Kaushal and Parthasarathi, J. Phys.{\bf A}:math \& Gen. {\bf 35} (2002) 8747; Parthasarathi, D. Parashar and R. S. Kaushal, J. Phys. {\bf A}:Math \& Gen. {\bf 37} (2004) 781, and the references therein.\\

[14]. A.L. Xavier,Jr. and M.A.M. Aguiar, Ann. Phys.(NY) {\bf 252} (1996) 458.\\

[15]. R.S. Kaushal and H.J. Korsch, Phys. Lett. {\bf A276} (2000) 47.\\

[16]. R.S. Kaushal, Int. Jour. Theo. Phys. {\bf 40} (2001) 835; R.S.Kaushal, Mod. Phys. Lett.{\bf A15} (2000) 1391.\\

[17]. V.P. Ermakov, Univ. Izv. Kiev {\bf 20} (1880),Series III, 1 ; H.R. Lewis, J. Math. Phys. {\bf 9} (1968) 1976.\\

[18]. I.S. Gradshteyn and I.M. Ryzhik, {\sf Table of Integrals, Series and Products} (Academic Press, INC., London,1980).\\

[19]. R. S. Kaushal, {\sf Classical and Quantum Mechanics of Noncentral Potentials: A Survey of Two-Dimensional Systems}, (Narosa Publishing House, New Delhi, Spriger-Verlag, Heidelberg, 1998) Chapters 2 and 3.\\

[20]. P.G. Drazin and R.S. Johnson, {\sf Soliton: An Introduction}, (Cambridge University Press, Cambridge,1990), Also see, R.S. Kaushal and D. Parashar, Ref.(8) above, Chapter 8.\\

[21]. See, for example, R.S. Kaushal, Class.\& Quant. Grav. {\bf15} (1998) 197; R.M. Hawkins and J.E. Lidsey, Phys. Rev. {\bf D66} (2002) 023523; F. L. Williams and P.G. Kevrekidis, Class. \& Quant. Grav. {\bf 20} (2003) L177.\\

[22]. See, for example, J.R. Ray and J. L. Reid, J. Math. Phys. {\bf 20} (1979) 2054, and Ref.(19), Chapter 7.\\ 

[23]. See, for example, V. S. Vladimirov, {\sf Methods of the Theory of functions of Many Complex Variables}, (MIT Press, Cambridge, MA., 1966).\\


\end{document}